\input{aipcheck}
\documentclass[final]{aipproc}
\layoutstyle{6x9}

\usepackage{amsmath}

\begin{document}

\title{ 
Chiral symmetry breaking beyond BCS
and theorem on the width of scalars
}

\author{P. Bicudo} {
address=Dep. F\'{\i}sica and CFTP Instituto Superior T\'ecnico
%Av. Rovisco Pais, 1049-001 Lisboa, Portugal
%\email{bicudo@ist.utl.pt}
}

%\affiliation{Dep. F\'{\i}sica and CFIF, Instituto Superior T\'ecnico,
%Av. Rovisco Pais, 1049-001 Lisboa, Portugal}

\begin{abstract}
We review chiral symmetry breaking at the BCS level, in the framework
of chiral invariant quark models and in the Schwinger-Dyson formalism.
We revisit the $\pi$ mass problem beyond the BCS level.
We show a theorem on the masses, on the widths and on the $q \bar q$ 
content of the scalar mesons $\sigma$ and $f_0$. 
\end{abstract}


\maketitle

%%%%%%%%%%%%%%%%%%%%%%%%%%%%%%%%%%%%%%%%%%%%%%%%%%%%%%%%%%%%%%%%%%%%%%%
%%                                                                   
%%
%%                                   SSS                             
%%
%%                                  S   S                            
%%
%%                                   S                               
%%
%%                                    S                              
%%
%%                                     S                             
%%
%%                                  S   S                            
%%
%%                                   SSS                             
%%
%%                                                                   
%%
%%%%%%%%%%%%%%%%%%%%%%%%%%%%%%%%%%%%%%%%%%%%%%%%%%%%%%%%%%%%%%%%%%%%%%%

\section{Introduction: chiral Symmetry Breaking at BCS}
\label{BCS}

With no quark loop, at the BCS level, equivalent to the rainbow/ladder truncation several
hadronic properties can be derived in the quark model with many-body techniques
pioneered by Scadron, and others
\cite{Nambu,Delbourgo,Yaouanc,Bicudo_thesis,Llanes-Estrada_thesis} ,
\\ - equivalent to the Schwinger-Dyson approach,
\\ - and in the fermion perspective, equivalent to quenched Lattice QCD.

In this talk the equations are written in terms of Feynman diagrams, for simplicity.
The propagators are quark propagators. The potential is not detailed since diferent
quark-quark interactions are used in the litterature. The interaction is only required
\\ - to be chiral invariant, an this is sufficient to reproduce qualitatively the phenomena
of low energy hadronic physics,
\\ - to account quantitavely for the observables, say $M_\pi$, $f_\pi$, $M_\rho, \cdots$.

At this BCS level we can start by computing the quark mass, solving the mass gap equation,
\begin{equation}
\includegraphics[width=0.7\textwidth]{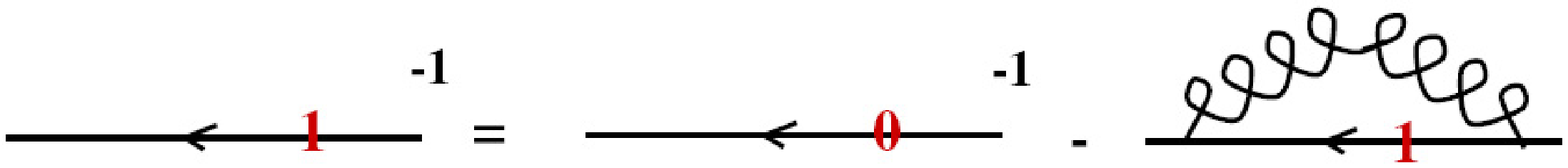} \ ,
\end{equation}
where $0$ labels the free quark propagator and $1$ labels the quark propagator at the BCS level, 
or ladder/rainbow truncation, in the series of Feynman diagrams.

Then, at the BCS level, the bound state equation is obtained summing the ladder,
\begin{equation}
\includegraphics[width=0.7\textwidth]{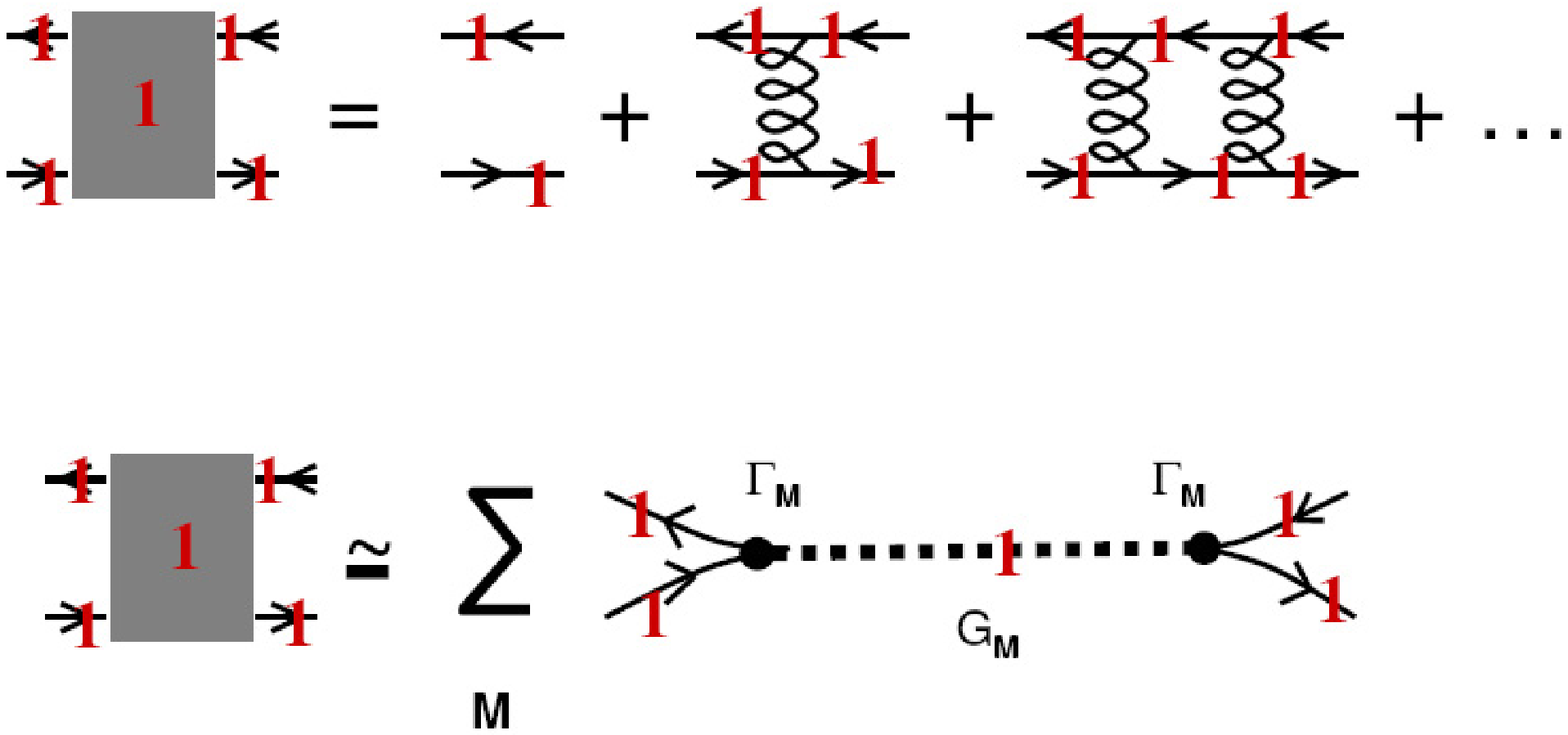} \ .
\end{equation}

For the vertex we get the Bethe-Salpeter equation,
\begin{equation}
\includegraphics[width=0.7\textwidth]{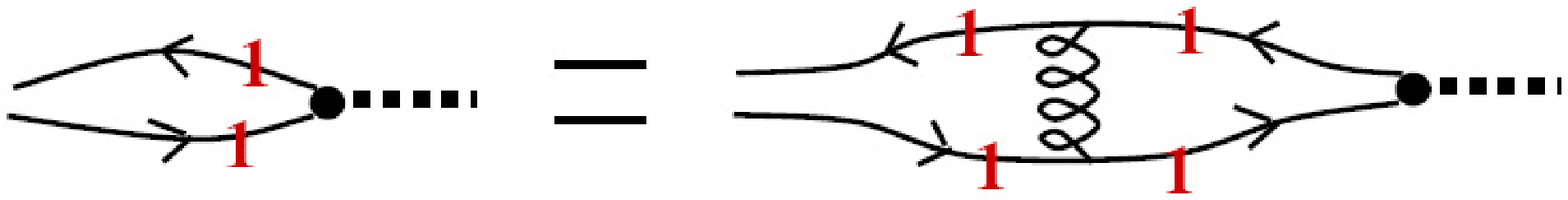} \ ,
\end{equation}
which in the equal-time approximation is 
equivalent to the Salpeter or RPA equation.
In any case onde obtains the Gell-Mann-Oakes-Renner relations,
implying that
$M_\pi \rightarrow 0$ when $m_c \rightarrow 0$; 
and one also gets all the hadron spectrum,
correct up to the quality of the potential,
\begin{equation}
\includegraphics[width=0.2\textwidth]{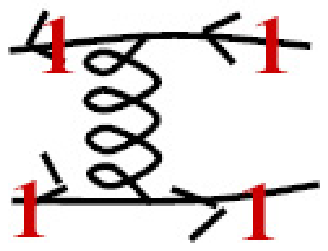} \ .
\end{equation}

The $\pi - \pi$ scattering can also be studied at the BCS level.
The sum of all consistent series of ladder with one-loop diagrams
sum was first performed by Hiller {\em et al.}, and
by Boroniowsky {\em et al.}
\cite{Hiller,Broniowsky}
in the NJL model.
Then Bicudo, Llanes-Estrada, Cotanch {\em et al.} 
\cite{Bicudo_PCAC,weall_pipi,Bicudo_piN,Llanes-Estrada_l1l2}
summed the same class of diagrams both in the
quark model and in the Schwinger-Dyson formalism,
\begin{equation}
\includegraphics[width=0.3\textwidth]{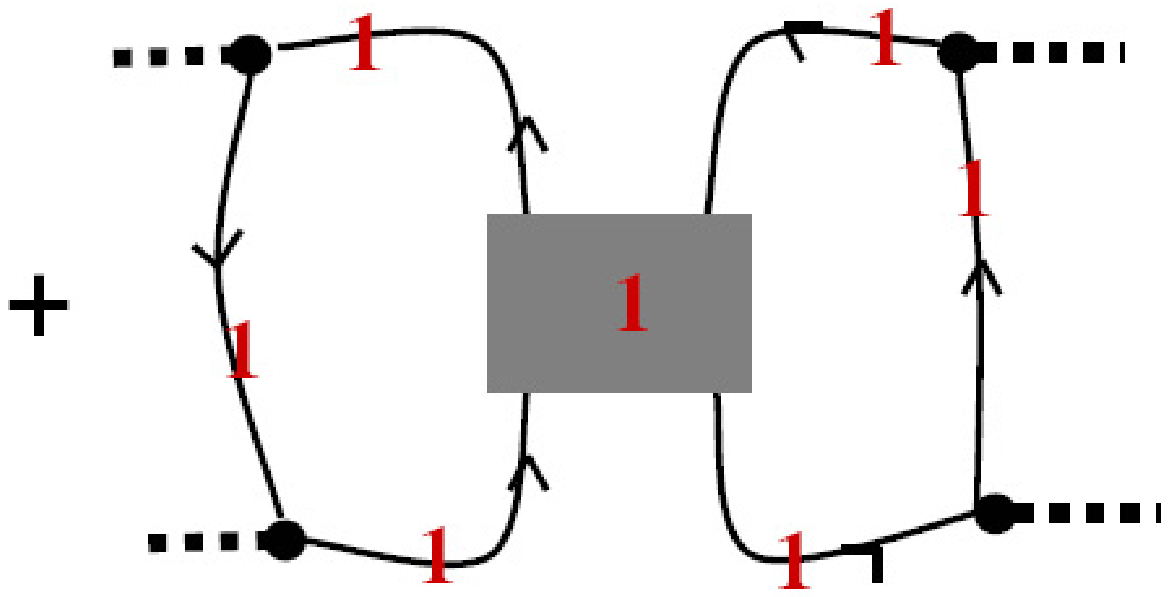}
\includegraphics[width=0.3\textwidth]{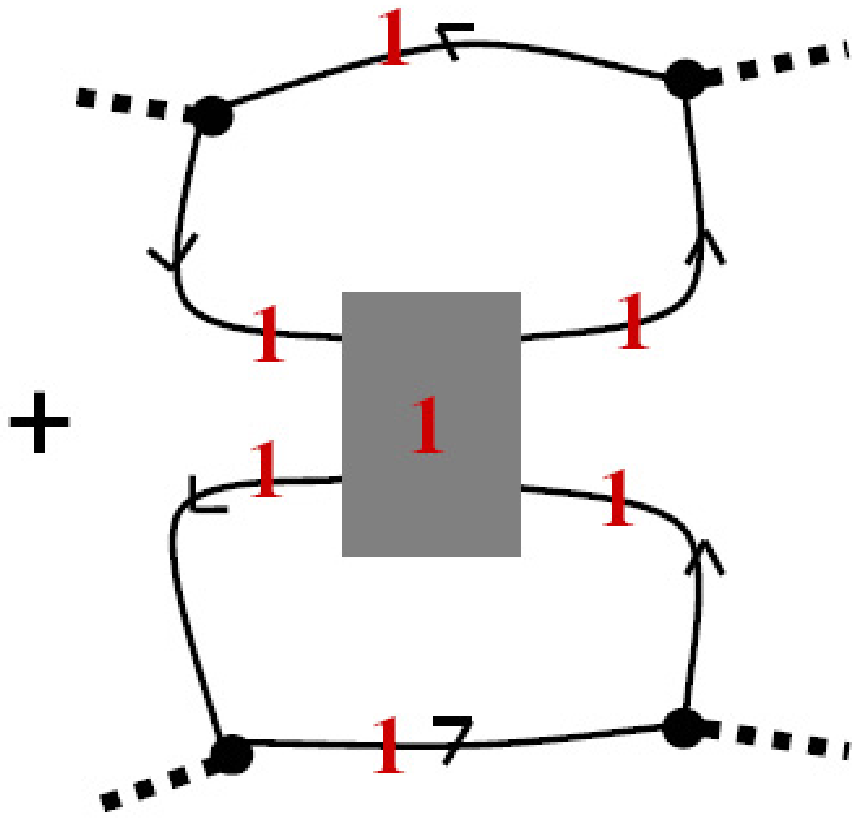}
\includegraphics[width=0.3\textwidth]{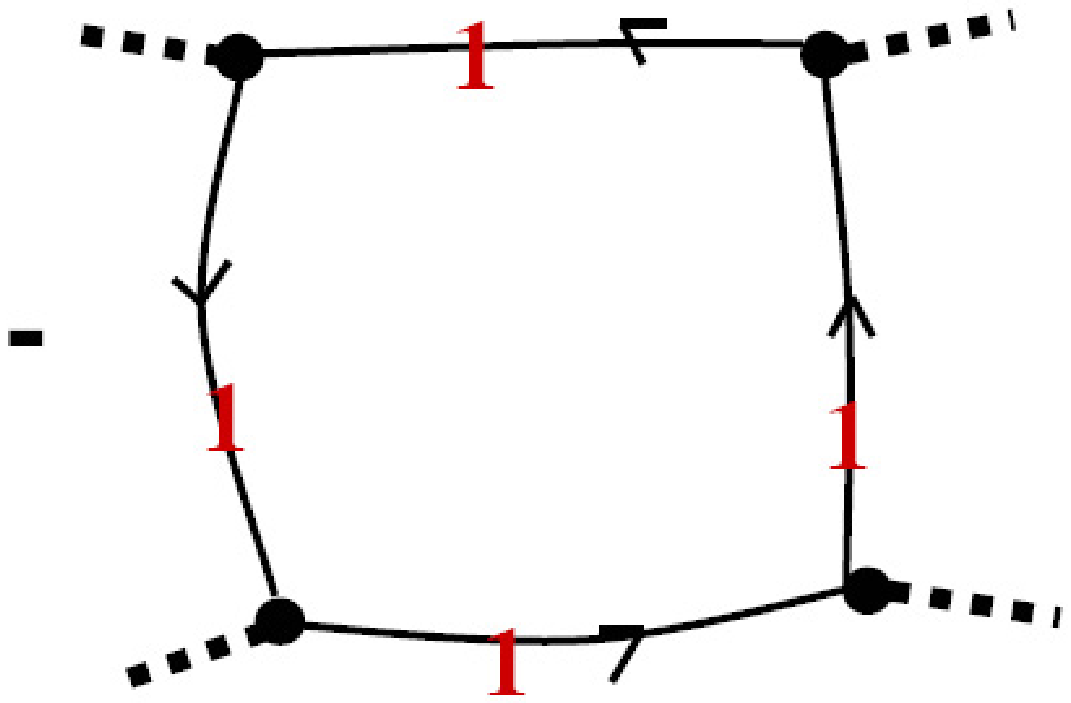}
\end{equation}
and this indeed reproduces the Chiral theorems, say the Adler Zero 
and the Weinberg Theorem.

\section{ Revisiting the $\pi$ mass problem beyond BCS}
\label{bBCS}

The BCS, or rainbow/ladder truncation, does not include coupled channels, nor decay
widths of mesons. A correct description of hadronic physics requires the inclusion of
coupled channel diagrams in the BS equation.

In effective effective field theories, without quarks or gluons, at least meson loops 
should be included,
\begin{equation}
\includegraphics[width=0.7\textwidth]{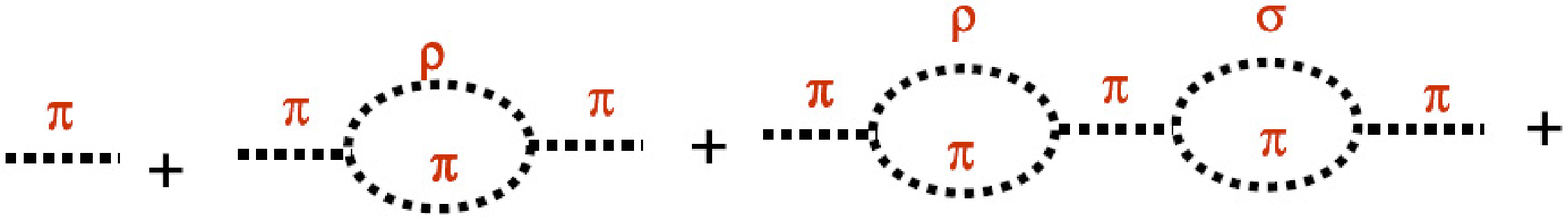}
\end{equation}

Applying this to the pion, it would produce a negative shift of ${M_\pi}^2$,
or imaginary shift of $M_\pi$. Thus this problem must be solved self-consistently.

At the quark level we must suplement the Bethe Salpeter kernel for the pion
(Isospin $I=1$) with the diagrams,
\begin{equation}
\includegraphics[width=0.6\textwidth]{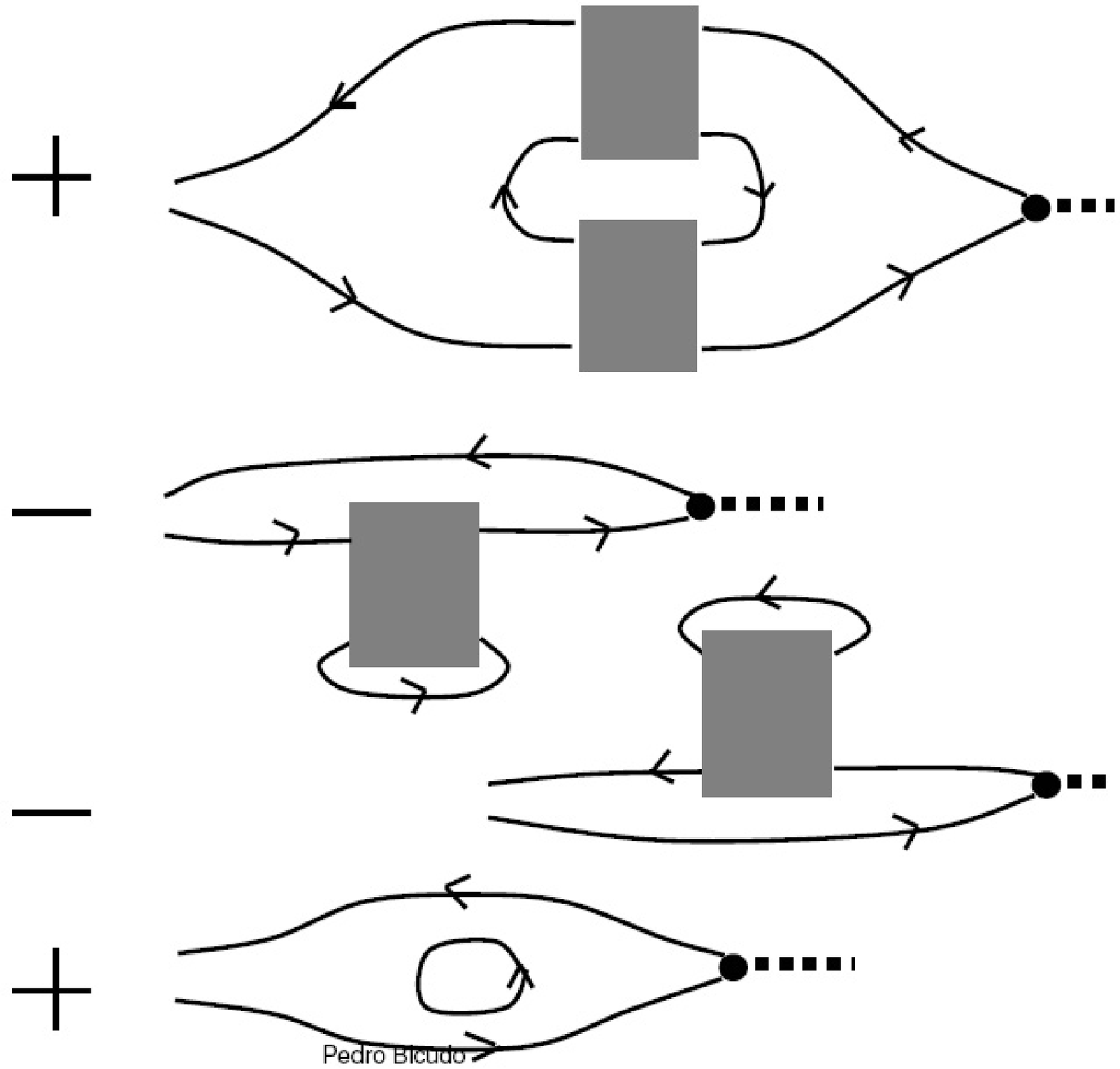} \ .
\end{equation}

Interestingly, this may account for
the $\eta'$ , $\eta$,  $\pi_0$ mass differences,
suplementing the isospin $I=1$ diagrams with the 
diagrams that only contribute to the $I=0$ mesons,
as in Lakhina and Bicudo
\cite{Lakhina}, 
with the diagrams,
\begin{equation}
\includegraphics[width=0.1\textwidth]{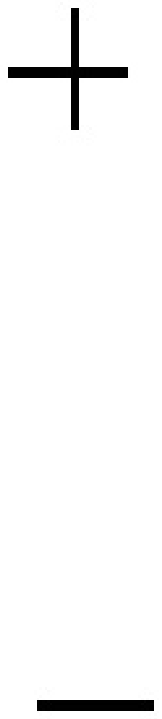} 
\includegraphics[width=0.5\textwidth]{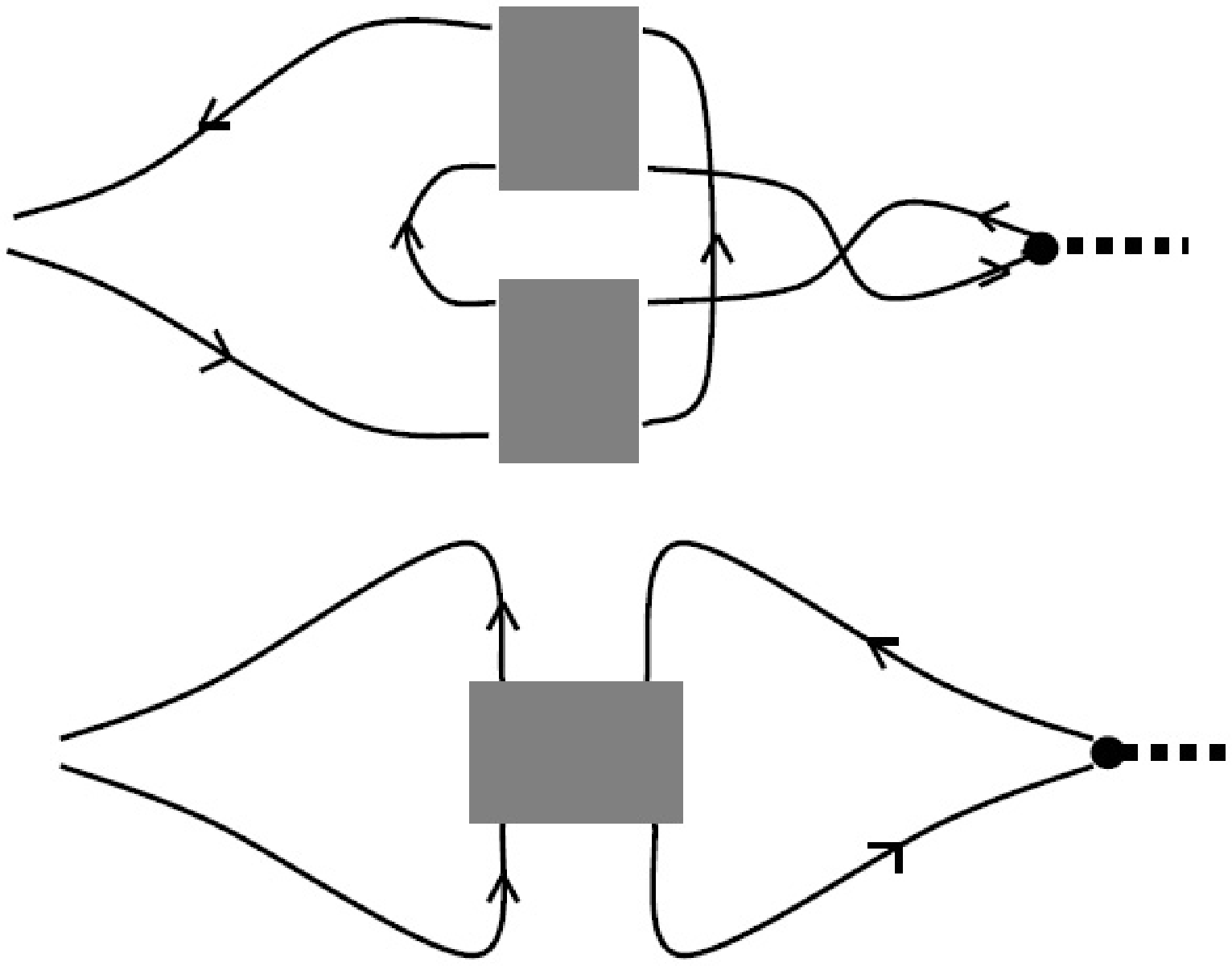} \ .
\end{equation}

\section{ A theorem on scalar $\sigma$, $f_0 \cdots$ widths and $\langle q \bar q \rangle$ content}
\label{Theorem}

To again solve the $M_\pi$ imaginary mass problem, now beyobd the BCS level, 
we have to return to the mass gap equation,
going beyond BCS, now with at least one extra quark loop.
\begin{equation}
\includegraphics[width=0.7\textwidth]{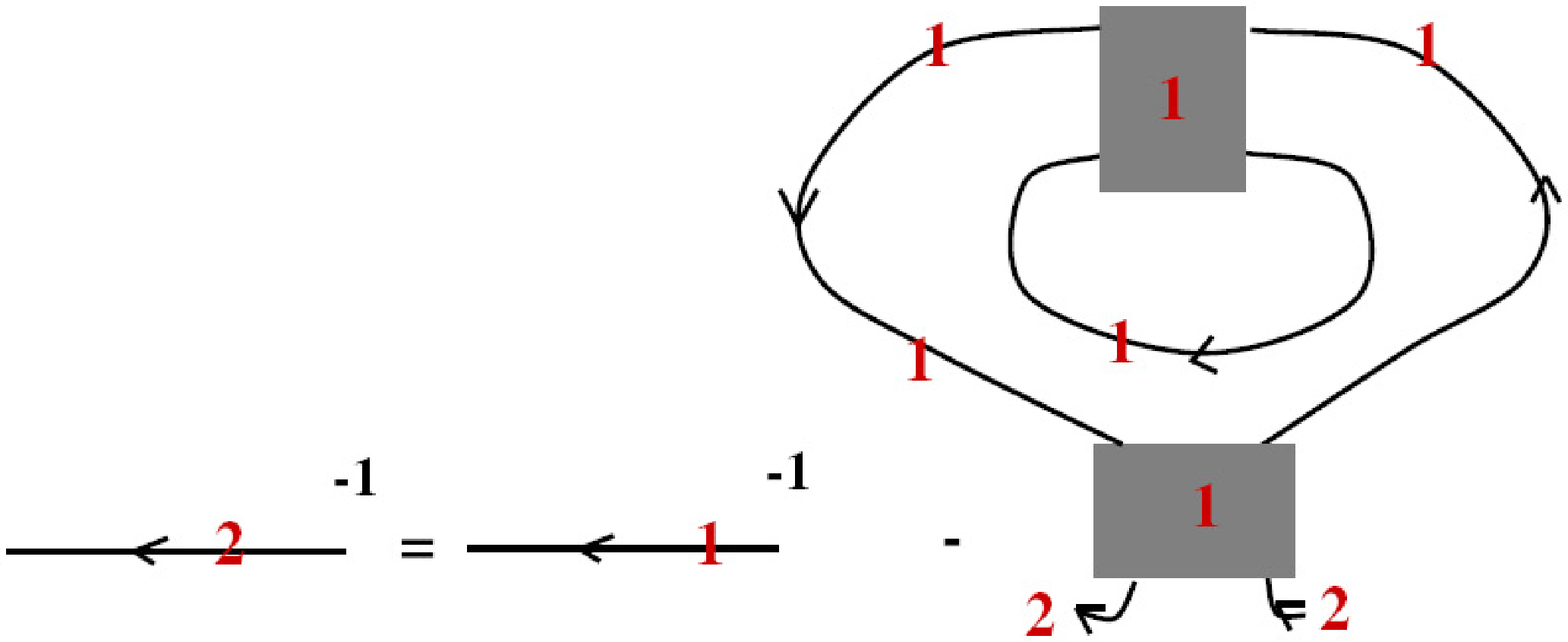}
\end{equation}
where $2$ labels the quark propagator beyond BCS, at 1- loop level.

This contribution to the mass gap equation was computed by Bicudo
\cite{Bicudo2},
see also Wojciech Broniowsky {\em et al}
\cite{Broniowsky}.

\vspace{1cm}
Importantly, the new beyond BCS term suppresses chiral symetry breaking,
like in the $\sigma$ model the $\phi^4$ term counteracts the $-\phi^2$ term,
\begin{equation}
\includegraphics[width=0.6\textwidth]{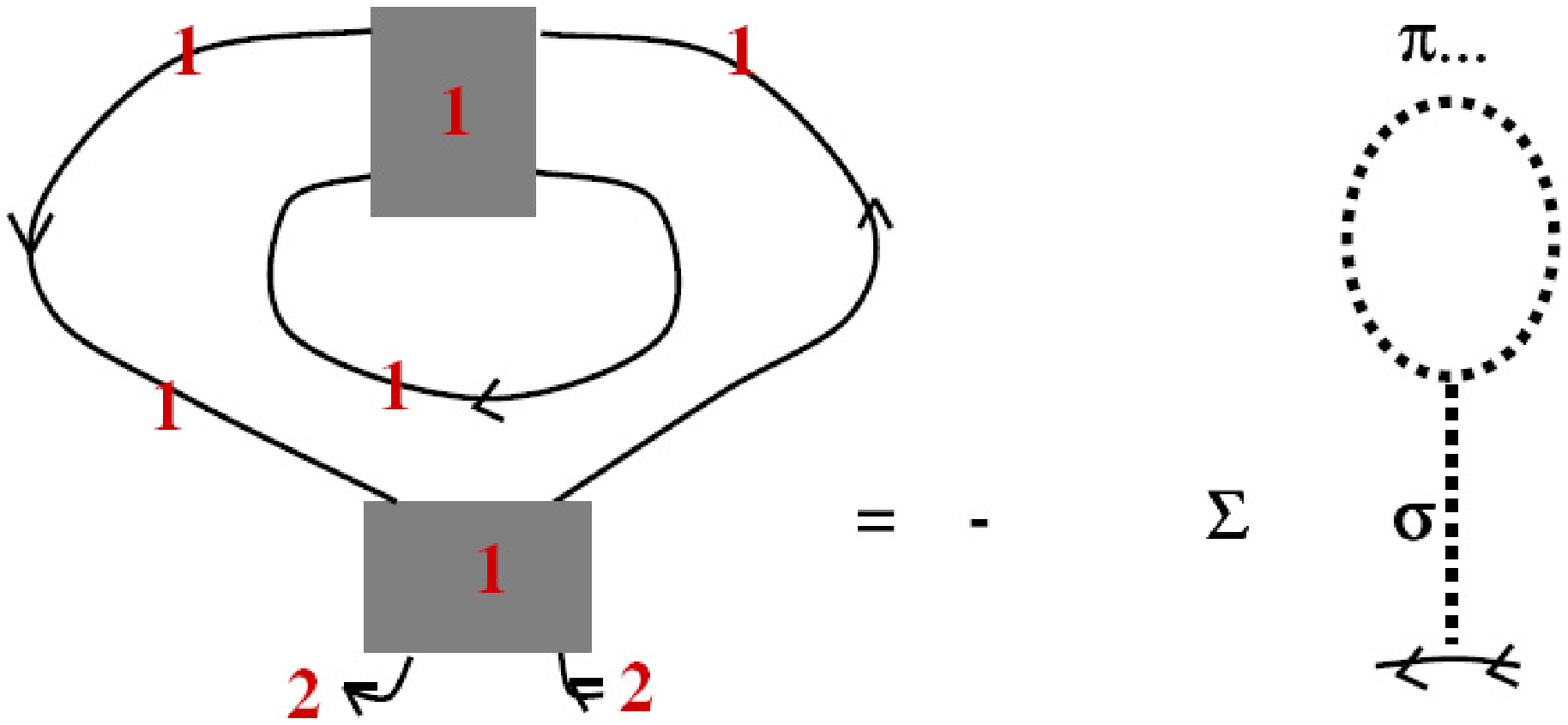}
\end{equation}
and this term must necessarily be smaller than the BCS level terms, to
preserve the spontaneous breaking of chiral symmetry.

Notice that this term increases with the decay width of the isospin $I=0$
scalars with a large $\langle q \bar q \rangle$ component, and with the mass of these
same $I=0$ scalars. Thus we arrive at 
a theorem stating that, 
\\ - either the isospin $I=0$ scalars have a small decay
\\ - or they have a large mass
\\ - or they have a small $\langle q \bar q \rangle$ component.

\vspace{1cm}
\section{Summary}

To conclude we find that,
\\ - either the $q \bar q$ coupling to the scalar meson is small,
\\ - or the scalar meson mass $M$ is large,
\\ - or the $\pi \pi$ coupling to the scalar meson is small.

Moreover the chiral invariant quark model or equivalently the Schwinger-Dyson formalism
reproduce several chirla theorems, and also reproduces the $U_A(1)$ splitings of the 
$M_{\eta'}$ , $M_\eta$, $M_{\pi_0}$ masses.

The results in a particular confining and chiral invariant quark model 
\cite{Bicudo2} are,
\\ - this one loop beyond BCS rainbow/ladder truncation decreases the quark condensate
$\langle q \bar q \rangle $ by 5\% to 50\%.
\\ - the lowest $q \bar q$  scalar has a mass $M \simeq 1$ GeV, noticing however that
the spin-orbit interaction is too strong and thus a better model might produce a larger
mass, 
\\ - the lowest $q \bar q$ scalar has a width $\Gamma$ between 50 MeV and 100 MeV.

\vspace{1cm}
%bbbbbbbbbbbbbbbbbbbbbbbbbbbbbbbbbbbbbbbbbbbbbbbbbbbbbbbbbbbb
%bbbbbbbbbbbbbbbbbbbbbbbbbbbbbbbbbbbbbbbbbbbbbbbbbbbbbbbbbbbb
%bb
%bb
%bbbbbbbbbbbbbbbbbbbbbbbbbbbbbbbbbbbbbbbbbbbbbbbbbbbbbbbbbbbb
%bbbbbbbbbbbbbbbbbbbbbbbbbbbbbbbbbbbbbbbbbbbbbbbbbbbbbbbbbbbb

\end{document}